\documentstyle[11pt,twoside]{article}
\def\greaterthansquiggle{\raise.3ex\hbox{$>$\kern-.75em\lower1ex\hbox{$\sim$}}}
\def\lessthansquiggle{\raise.3ex\hbox{$<$\kern-.75em\lower1ex\hbox{$\sim$}}}
\newcommand{\beq}{\begin{equation}}
\newcommand{\eeq}{\end{equation}}
\newcommand{\beqa}{\begin{eqnarray}}
\newcommand{\eeqa}{\end{eqnarray}}
\newcommand{\beqan}{\begin{eqnarray*}}
\newcommand{\eeqan}{\end{eqnarray*}}
\newcommand{\ba}{\begin{array}}
\newcommand{\ea}{\end{array}}
\newcommand{\no}{\nonumber}

\newcommand{\lets}{\lessthansquiggle}

\newcommand{\ra}{\rightarrow}

\newcommand{\ve}{\varepsilon}

\newcommand{\dg}{\dagger}

\newcommand{\wh}{\widehat}

\newcommand{\cL}{{\cal L}}
\newcommand{\M}{{\cal M}}

\newcommand{\cO}{{\cal O}}

\newcommand{\Q}{{\cal Q}}

\newcommand{\dfrac}{\displaystyle \frac}

\def\nz{\ifmmode {I\hskip -3pt N} \else {\hbox {$I\hskip -3pt N$}}\fi}
\def\zz{\ifmmode {Z\hskip -4.8pt Z} \else
       {\hbox {$Z\hskip -4.8pt Z$}}\fi}
\def\qz{\ifmmode {Q\hskip -5.0pt\vrule height6.0pt depth 0pt
       \hskip 6pt} \else {\hbox
       {$Q\hskip -5.0pt\vrule height6.0pt depth 0pt\hskip 6pt$}}\fi}
\def\rz{\ifmmode {I\hskip -3pt R} \else {\hbox {$I\hskip -3pt R$}}\fi}
\def\cz{\ifmmode {C\hskip -4.8pt\vrule height5.8pt\hskip 6.3pt} \else
       {\hbox {$C\hskip -4.8pt\vrule height5.8pt\hskip 6.3pt$}}\fi}

\def\au{{\setbox0=\hbox{\lower1.36775ex%
\hbox{''}\kern-.05em}\dp0=.36775ex\hskip0pt\box0}}
\def\ao{{}\kern-.10em\hbox{``}}

\normalsize
\begin{document}
\bibliographystyle{plain}
\begin{titlepage}
\begin{flushright}
UWThPh-1995-18\\
June 28, 1995
\end{flushright}
\vspace*{4cm}
\begin{center}
{\Large \bf
The Electromagnetic Interaction in \\
Chiral Perturbation Theory*}\\[50pt]
H. Neufeld$^\diamond$ and H. Rupertsberger$^{\diamond\diamond}$  \\
Institut f\"ur Theoretische Physik der Universit\"at Wien\\
Boltzmanngasse 5, A-1090 Wien, Austria\\[25pt]
$^\diamond$ neufeld@ariel.pap.univie.ac.at\\
$^{\diamond\diamond}$ rupert@pap.univie.ac.at\\
\vfill
{\bf Abstract} \\
\end{center}
\noindent
We investigate electromagnetic effects in the framework of chiral
perturbation theory. Using a completely independent method, we
confirm Urech's results for the divergences of the one--loop
functional in the electromagnetic sector. We perform a one--loop
analysis of all $P_{\ell 2}$ ($P = \pi, K, \eta$) and the
$K_{\ell 3}$ form factors
$f_+^{K^+\pi^0}(0)$, $f_+^{K^0\pi^-}(0)$, including a systematic
treatment of the $\cO(e^2p^2)$ contributions in the mesonic part.
We illustrate our results by several numerical estimates.

\vfill
\noindent * Work supported in part by Jubil\"aumsfonds der
Oesterreichischen Nationalbank, Project No. 5051.
\end{titlepage}

\section{Introduction}
\label{sec: Introduction}
\renewcommand{\theequation}{\arabic{section}.\arabic{equation}}
\setcounter{equation}{0}

In the standard model of strong and electroweak interactions, the
violation of the isospin symmetry has two different origins. First of
all, it can be traced back to the mass difference of up
and down quark. Secondly, also electromagnetism
induces isospin breaking effects.

In the confinement region of the standard model, the usual
perturbative methods are not applicable. In order to obtain testable
theoretical predictions also in this case one has to resort to a
so--called low--energy effective theory. With an appropriately chosen
effective Lagrangian, chiral perturbation
theory \cite{Weinberg,GL1,GL2} (which is just the effective field theory of
the standard model at low energies) is mathematically equivalent
\cite{Leutwyler} to the underlying fundamental theory. Therefore, chiral
perturbation theory presents the natural framework for the discussion
of isospin breaking effects in the low energy range.

Isospin violating contributions related to $m_u \not= m_d$ are well under
control from the theoretical point of view. They are fully
described by the effective field theory of the strong interactions. Up to
the chiral order $p^4$, the associated low--energy constants
\cite{GL2} have been determined \cite{DAPHNE2} with rather good accuracy.

In principle, it is also straightforward to establish the
theoretical framework for the description of electromagnetic
effects. First of all, the photon field has to be included as an
additional dynamical degree of freedom. Then one has to
construct the most general Lagrangian of the desired order $e^2
p^{2n}$ respecting all the symmetries of the standard model. To
lowest electromagnetic order $e^2 p^0$, only a single term appears
\cite{EGPR}. But already at the next--to--leading order $e^2 p^2$,
there are 14 linear independent terms \cite{Urech,NeuRup}
entering the effective Lagrangian . The associated coupling
constants $K_i$ absorb the divergences generated by one--loop
graphs with a virtual photon or a vertex from the Lagrangian of
$\cO(e^2 p^0)$. The divergent parts of the couplings $K_i$ have
been determined in Ref. \cite{Urech}. However, the finite parts
$K_i^r$ of the electromagnetic low--energy constants are
remaining as free parameters. At this point one encounters the
main difference between the strong and the electromagnetic
sector. In contrast to the low--energy constants of the strong
interactions, only rough order of magnitude estimates for the
$K_i^r$ are presently available \cite{Urech,NeuRup}.

With the methods sketched above, it is possible to obtain the
formal expressions of the electromagnetic contributions to
$\cO(e^2 p^2)$ for any mesonic observable. So far, only a
small number of applications \cite{Urech,NeuRup} of this kind
has been worked out. It is one of the purposes of the present
paper to add some new results to this list.

In Sect. \ref{sec: Electromagnetism}, we briefly review the
construction of the electromagnetic effective Lagrangian. The
one-loop renormalization in the electromagnetic sector is
discussed in Sect. \ref{sec: Renormalization}. There, we also
give an alternative determination of eight linear combinations
of the renormalization constants $K_i^{\rm div}$ which serves as
an independent test of the general results obtained in Ref.
\cite{Urech}. In Sect. \ref{sec: Applications} we illustrate
the power and the limits of simple order of magnitude estimates
for the $K_i^r$ in the mass spectrum of the pseudoscalar mesons.
A complete list of the $P_{\ell 2}$ form factors including the
(mesonic) electromagnetic contributions of $\cO(e^2 p^2)$ is
presented in Sect. \ref{sec: Pl2}. The analogous expressions for
the $K_{\ell 3}$ form factors $f_+^{K^+\pi^0}(0)$ and $f_+^{K^0
\pi^-}(0)$ are given in Sect. \ref{sec: Kl3}. In both cases, our
results are illustrated by numerical estimates discriminating
the pure QCD contributions and the electromagneting ones for
certain isospin violating quantities. Finally, our conclusions
are summarized in Sect. \ref{sec: Conclusions}.

\section{The Effective Chiral Lagrangian of Electromagnetism}
\label{sec: Electromagnetism}
\renewcommand{\theequation}{\arabic{section}.\arabic{equation}}
\setcounter{equation}{0}

Chiral perturbation theory \cite{Weinberg,GL1,GL2} permits a systematic
low--energy expansion of the generating functional $Z[v,a,s,p]$ of QCD. This
quantity is defined in terms of the vacuum--to--vacuum amplitude
\beq
e^{iZ[v,a,s,p]} = \langle 0\mbox{ out}|0\mbox{ in}\rangle_{v,a,s,p}
\eeq
associated with the Lagrangian
\beq
\cL = \cL^0_{\rm QCD} + \bar q \gamma^\mu(v_\mu + a_\mu \gamma_5)q
- \bar q (s - ip\gamma_5)q. \label{LQCD}
\eeq
$\cL_{\rm QCD}^0$ is the QCD Lagrangian with the masses of the three light
quarks $q = (q_u,q_d,q_s)^T$ set to zero. $v_{\mu}, a_{\mu}, s, p$ are external
sources represented by hermitian $3 \times 3$ matrices in flavour space. The
Green functions of the vector, axial--vector, scalar and pseudoscalar quark
currents can then be obtained by evaluating the functional derivatives of
$Z[v,a,s,p]$ at $v=a=p=0$, $s=\M_{quark}=\mbox{ diag}(m_u,m_d,m_s)$.

The effective chiral Lagrangian of QCD consists of a string of terms
\beq
\cL_{\rm eff} = \cL_2 + \cL_4 + \cL_6 + ... \enspace ,
\eeq
organized in powers of momenta and meson masses, respectively. The
lowest order term $\cL_2$ is the nonlinear sigma model Lagrangian in
the presence of external fields\footnote{Our notation is the same as
in Refs. \cite{EGPR,NeuRup}.}:
\beq
\cL_2 = \frac{F^2}{4} \langle u_\mu u^\mu + \chi_+\rangle. \label{L2}
\eeq

The generating functional $Z[v,a,s,p]$ is
given by the expansion of the effective meson field theory in the
number of loops,
\beq
Z = Z_2 + Z_4 + Z_6 + ... \enspace .
\eeq
The leading term coincides with the classical action associated with
$\cL_2$.

At next--to--leading order $p^4$, the generating functional consists of
the following terms: one--loop graphs generated by the vertices of
$\cL_2$, tree graphs involving one vertex from $\cL_4$ and finally a
contribution to account for the chiral anomaly.

Also electromagnetic
processes where only external photon fields $A_{\mu}$ are present can be
treated within this framework. One simply performs the substitution
\beq
v_{\mu} = - e Q A_{\mu},
\eeq
where
\beq
Q = \frac{1}{3} \mbox{ diag}(2,-1,-1)
\eeq
is the electromagnetic charge matrix.

In those cases where virtual photons are involved, the above approach is, of
course, not sufficient any more. Now the photon field has to be included as an
additional dynamical degree of freedom. In order to construct the pertinent
effective Lagrangian of electromagnetic order $e^2$, one introduces spurion
fields $\Q_{L,R}(x)$ \cite{NeuRup} transforming  as
\beqa
\Q_{L,R} &\stackrel{G}{\ra}& h(\pi) \Q_{L,R} h(\pi)^\dg, \no \\
\Q_{L,R} &\stackrel{P}{\ra}& \Q_{R,L},
\eeqa
under the chiral group $G = SU(3)_L \times SU(3)_R$ and parity
$P$, respectively. The nonlinear realization $h(\pi)$ of $G$
\cite{CCWZ} is defined by the action of the chiral group $G$ on
the coset space $C = SU(3)_L \times SU(3)_R/SU(3)_V$:
\beqa
u(\pi) &\stackrel{G}{\ra}& g_R u(\pi) h(\pi)^\dg = h(\pi) u(\pi)g_L^\dg,
\no \\
u(\pi) &\in& C, \no \\
g_{L,R} &\in& SU(3)_{L,R}.
\eeqa
The Goldstone fields $\pi_i$ $(i = 1,\ldots,8)$ are
coordinates of the coset space $C$. We use the parametrization
$$
u = \exp (i \Phi/\sqrt{2}\; F),
$$

\beq
\Phi = \Phi^\dg = \left( \ba{ccc}
\dfrac{\pi_3}{\sqrt{2}} + \dfrac{\pi_8}{\sqrt{6}} & \pi^+ & K^+ \\[10pt]
\pi^- & - \dfrac{\pi_3}{\sqrt{2}} + \dfrac{\pi_8}{\sqrt{6}} & K^0 \\[10pt]
K^- & \bar K^0 & - \dfrac{2 \pi_8}{\sqrt{6}}
\ea \right). \label{conv}
\eeq

Alternatively, one can also define \cite{EGPR} spurions
$Q_{L,R}$ with the transformation properties
\beq
\ba{l}
Q_L \stackrel{G}{\ra} g_L Q_L g_L^\dg, \quad
Q_R \stackrel{G}{\ra} g_R Q_R g_R^\dg.
\ea
\eeq
The $\Q_{L,R}$ are related to $Q_{L,R}$ by
\beqa
\Q_L &=& u Q_L u^\dg, \no \\
\Q_R &=& u^\dg Q_R u. \label{DEF}
\eeqa
At the end $Q_{L,R}$ will be identified with the charge matrix $Q$.

To lowest order $e^2 p^0$, the electromagnetic effective Lagrangian
contains a single term \cite{EGPR}
\beq
\left. \cL\right|_{\cO(e^2p^0)} = F^4 e^2 Z \; \langle \Q_L \Q_R\rangle,
\label{LELM}
\eeq
with a real and dimensionless coupling constant $Z$. The effective
Lagrangians (\ref{L2}) and (\ref{LELM}) generate the lowest--order
contributions to the masses of the pseudoscalar mesons from QCD and the
electromagnetic interaction, respectively:
\beqa
\wh M^2_{\pi^\pm} &=& 2B \wh m + 2 e^2 Z F^2, \no \\
\wh M^2_{\pi^0} &=& 2B \wh m , \no \\
\wh M^2_{K^\pm} &=& B\left[ (m_s + \wh m) - \frac{2\ve}{\sqrt{3}}
(m_s - \wh m)\right] + 2e^2 Z F^2, \no \\
\wh M^2_{\stackrel{(-)}{K}{}^0} &=& B \left[(m_s + \wh m) +
\frac{2\ve}{\sqrt{3}} (m_s - \wh m)\right] , \no \\
\wh M^2_\eta &=& \frac{4}{3} B\left( m_s + \frac{\wh m}{2}\right),
\label{treemass}
\eeqa
where $\wh m$ denotes the mean value of the light quark masses,
\beq
\wh m = \frac{1}{2} (m_u + m_d),
\eeq
and $B$ is the vacuum condensate parameter contained in $\chi_+$.
The mixing angle
\beq
\ve = \frac{\sqrt{3}}{4} \; \frac{m_d - m_u}{m_s - \wh m} \label{epsilon}
\eeq
relates $\pi_3$, $\pi_8$ to the (tree--level) mass eigenfields
$\wh \pi_0$, $\wh \eta$:
\beqa
\pi_3 &=& \wh \pi^0 - \ve \wh \eta, \no \\
\pi_8 &=& \ve \wh \pi^0 + \wh \eta .
\eeqa
Terms of higher than linear order in $\ve$ have been neglected. In
accordance with Dashen's theorem \cite{Dashen}, the lowest order
electromagnetic Lagrangian (\ref{LELM}) contributes an equal amount to the
squared masses of $\pi^\pm$, $K^\pm$. It does not contribute to the masses of
$\pi^0$, $K^0$, $\bar K^0$ or $\eta$, nor does it generate
$\pi^0$--$\eta$ mixing.
The relation
\beq
M^2_{\pi^\pm} - M^2_{\pi^0} = 2 e^2 Z F^2 + \cO(e^2 p^2), \label{PIDIFF}
\eeq
resulting from (\ref{treemass}) implies $Z \simeq 0.8$ as numerical value.

At next--to--leading order $e^2p^2$ one finds the following list of local
counterterms \cite{Urech}:
\beqa
\left. \cL\right|_{\cO(e^2p^2)} &=&
F^2 e^2 \{ \frac{1}{2} K_1 \; \langle \Q^2_L + \Q^2_R\rangle \; \langle u_\mu
u^\mu\rangle \no \\
&& \mbox{} + K_2 \; \langle \Q_L \Q_R\rangle \; \langle u_\mu u^\mu
\rangle \no \\
&& \mbox{} - K_3 \; [\langle \Q_L u_\mu\rangle \; \langle \Q_L u^\mu
\rangle + \langle \Q_R u_\mu\rangle \; \langle \Q_R u^\mu\rangle ] \no \\
&& \mbox{} + K_4 \; \langle \Q_L u_\mu\rangle \; \langle \Q_R u^\mu \rangle
\no \\
&& \mbox{} + K_5 \; \langle(\Q^2_L + \Q^2_R) u_\mu u^\mu\rangle \no \\
&& \mbox{} + K_6 \; \langle (\Q_L \Q_R + \Q_R \Q_L) u_\mu u^\mu\rangle \no \\
&& \mbox{} + \frac{1}{2} K_7 \; \langle \Q_L^2 + \Q^2_R\rangle \; \langle
\chi_+\rangle \no \\
&& \mbox{} + K_8\; \langle \Q_L \Q_R\rangle \; \langle \chi_+\rangle \no \\
&& \mbox{}+ K_9 \; \langle (\Q_L^2 + \Q_R^2) \chi_+\rangle \no \\
&& \mbox{} + K_{10}\; \langle(\Q_L \Q_R + \Q_R \Q_L) \chi_+\rangle \no \\
&& \mbox{} - K_{11} \; \langle(\Q_L \Q_R - \Q_R \Q_L) \chi_-\rangle \no \\
&& \mbox{}- iK_{12}\; \langle(\wh \nabla_\mu \Q_L \Q_L -
\Q_L \wh \nabla_\mu \Q_L - \wh \nabla_\mu \Q_R \Q_R +
\Q_R \wh \nabla_\mu \Q_R) u^\mu\rangle \no \\
&& \mbox{}+ K_{13} \; \langle \wh \nabla_\mu \Q_L \wh \nabla^\mu \Q_R
\rangle \no \\
&& \mbox{} + K_{14} \; \langle \wh \nabla_\mu \Q_L \wh \nabla^\mu \Q_L +
\wh \nabla_\mu \Q_R \wh \nabla^\mu \Q_R\rangle  \}, \label{LE2P2}
\eeqa
where
\beqa
\wh \nabla_\mu \Q_L &=& \nabla_\mu \Q_L + \frac{i}{2} [u_\mu,\Q_L] =
u D_\mu Q_L u^\dg , \no \\
\wh \nabla_\mu \Q_R &=& \nabla_\mu \Q_R - \frac{i}{2} [u_\mu,\Q_R] =
u^\dg D_\mu Q_R u .
\eeqa
In order to obtain a linear independent set of terms in (\ref{LE2P2}), the
Cayley--Hamilton theorem,
\beq
P_A(A) \equiv 0, \label{P}
\eeq
has been used. The polynomial function $P_A$ is defined by
$P_A(\lambda) = \det(A - \lambda {\bf 1})$. Explicitly, the identity
(\ref{P}) reads:
\beq
- A^3 + \langle A\rangle A^2 + \frac{1}{2} (\langle A^2\rangle -
\langle A\rangle^2)A + \frac{1}{3} [\langle A^3\rangle -
\frac{3}{2} \langle A^2\rangle \; \langle A\rangle + \frac{1}{2}
\langle A\rangle^3] = 0. \label{CayHam}
\eeq
Replacing $A$ by $A \pm B$ in (\ref{CayHam}) yields identities which can
then be used to derive the relations
\beq
\langle \Q_I u_\mu \Q_I u^\mu\rangle = \frac{1}{2} \langle \Q_I^2\rangle\;
\langle u_\mu u^\mu\rangle - 2 \langle \Q_I^2 u_\mu u^\mu\rangle
+ \langle \Q_I u_\mu\rangle \; \langle \Q_I u^\mu\rangle,
\qquad I = L,R,
\eeq
and
\beq
\langle \Q_L u_\mu \Q_R u^\mu\rangle = \frac{1}{2} \langle \Q_L\Q_R\rangle\;
\langle u_\mu u^\mu\rangle -  \langle (\Q_L \Q_R + \Q_R \Q_L)
u_\mu u^\mu\rangle
+ \langle \Q_L u_\mu\rangle \; \langle \Q_R u^\mu\rangle.
\label{id}
\eeq

Furthermore,
the term $i \langle \Q_L^2 - \Q_R^2\rangle \langle \chi_-\rangle$
vanishes once $Q_L = Q_R = Q$ is inserted. The expression
$i \langle (\Q_L^2 - \Q_R^2)\chi_-\rangle$ does not contribute because
$[\M,Q] = 0$, and $i \langle \Q_L \Q_R\rangle \langle \chi_-\rangle$
is forbidden by P invariance. Finally, partial integration and the
equation of motion allows to relate
\beq
i \langle(\wh \nabla_\mu \Q_L \Q_R - \Q_R \wh \nabla_\mu \Q_L -
\wh \nabla_\mu \Q_R \Q_L + \Q_L \wh \nabla_\mu \Q_R)u^\mu\rangle
\label{rel1}
\eeq
to
\beqa
\langle  \Q_L \Q_R\rangle\;\langle u_\mu u^\mu\rangle - 3\langle(
\Q_L \Q_R + \Q_R \Q_L) u_\mu u^\mu \rangle
+ 2 \langle \Q_L u_\mu\rangle \; \langle \Q_R u^\mu\rangle + \no \\
\frac{1}{2} \langle (\Q_L \Q_R - \Q_R \Q_L) \chi_-\rangle +
\cO(e^2). \label{rel2}
\eeqa

\section{One--Loop Renormalization in the Electromagnetic Sector}
\label{sec: Renormalization}
\renewcommand{\theequation}{\arabic{section}.\arabic{equation}}
\setcounter{equation}{0}

One--loop graphs with a virtual photon or one vertex from (\ref{LELM}) are, in
general, divergent. These divergences associated with polynomial expressions
of order $e^2 p^2$ are absorbed by an
appropriate renormalization of the coupling constants in
(\ref{LE2P2}). To this end, the $K_i$ are decomposed in two parts:
\beq
K_i = K_i^r(\mu) + \Sigma_i \Lambda(\mu) . \label{renorm}
\eeq
The divergence is contained in the function $\Lambda(\mu)$. In dimensional
regularization, this scale dependent term is given by
\beq
\Lambda(\mu) = \frac{\mu^{d-4}}{(4\pi)^2} \left\{ \frac{1}{d-4} -
\frac{1}{2} [\ln (4\pi) + \Gamma'(1) + 1]\right\}.
\eeq
The renormalized electromagnetic low--energy constants $K_i^r(\mu)$ are, in
principle, measurable quantities. The constants $\Sigma_i$ govern the scale
dependence of the $K_i^r(\mu)$,
\beq
K_i^r(\mu_2) = K_i^r(\mu_1) + \frac{\Sigma_i}{(4\pi)^2}
\ln(\frac{\mu_1}{\mu_2}),
\eeq
and they also determine the so--called ``chiral logs''. In any physical
amplitude, the scale dependence always cancels between the loop and the
counterterm contributions containing the renormalized coupling constants.

A complete list of the renormalization constants $\Sigma_i$ has been worked
out in Ref. \cite{Urech} by evaluating the divergent part of the generating
functional. We have performed an independent check \cite{NeuRup} of the values
given there by evaluating the (potentially) divergent parts of several
observables. The requirement that the divergences associated with these
quantities should vanish produces a certain
number of conditions to be fulfilled by the $\Sigma_i$.
We have restricted our analysis to the masses of the
pseudoscalar mesons, the axial--vector decay constants $F_P$ and the
$P_{\ell 3}$ form
factors. In this case, only the following linear combinations of the
electromagnetic coupling constants appear:
\beq
\ba{lll}
S_1 = K_1 + K_2, & \qquad S_2 = K_5 + K_6, & \qquad
S_3 = -2K_3 + K_4 , \\[7pt]
S_4 = K_7 + K_8, & \qquad S_5 = K_9 + 2K_{10} + K_{11}, & \qquad
S_6 = K_8, \\[7pt]
S_7 = K_{10} + K_{11}, & \qquad S_8 = - K_{12}.
\ea
\eeq
In analogy to (\ref{renorm}), the associated renormalization constants
$\Delta_i$ are defined by
\beq
S_i = S_i^r(\mu) + \Delta_i \; \Lambda(\mu). \label{Si}
\eeq
The finiteness of the electromagnetic contributions to the
meson masses implies the relations
\beqa
\Delta_3 &=& - \frac{2}{3} \; \Delta_2 + 3Z , \no \\
\Delta_4 &=&  \Delta_1 + \frac{1}{3} \; \Delta_2 -\frac{1}{2} \; Z, \no \\
\Delta_5 &=& \frac{1}{6} \; \Delta_2 + \frac{3}{4} + \frac{11}{4} \; Z,
\no \\
\Delta_6 &=& Z, \no \\
\Delta_7 &=& \frac{1}{6} \; \Delta_2 + \frac{3}{4} + \frac{5}{4} \; Z.
\label{finite}
\eeqa
The analogous procedure for $F_{K^0}$ yields the relation
\beq
6 \Delta_1 + 2 \Delta_2 - 9 Z = 0. \label{RENFK0}
\eeq
Combined with the expression for $\Delta_3$, (\ref{RENFK0}) also
renormalizes the electromagnetic contributions to $F_{\pi^0}$ and
$F_\eta$. The requirement that $F_{\pi^\pm}$ (or $F_{K^\pm}$)
should be finite implies the relation
\beq
12\Delta_1 + 10 \Delta_2 - 18 \Delta_8 + 9 - 27Z = 0.
\label{RENFPIP}
\eeq
Finally, an inspection of the divergent terms in the $K_{\ell 3}$
form factor $f_+^{K^0\pi^-}(0)$ gives
\beq
\Delta_8 = - \frac{1}{4} . \label{DELTA8}
\eeq
This provides us with the necessary number of equations for the
determination of the eight renormalization constants
$\Delta_1, \ldots,\Delta_8$:
\beqa
\Delta_1 &=& \Sigma_1+\Sigma_2 \; = \; \dfrac{3}{4} + Z, \no \\
\Delta_2 &=& \Sigma_5+\Sigma_6 \; = \; - \dfrac{9}{4} + \dfrac{3}{2}Z, \no \\
\Delta_3 &=& -2\Sigma_3+\Sigma_4 \; = \; \dfrac{3}{2} + 2Z, \no \\
\Delta_4 &=& \Sigma_7+\Sigma_8 \; = \; Z, \no \\
\Delta_5 &=& \Sigma_9+2\Sigma_{10}+\Sigma_{11} \; = \; \dfrac{3}{8}+3Z,
\no \\
\Delta_6 &=& \Sigma_8 \; = \; Z, \no \\
\Delta_7 &=& \Sigma_{10}+\Sigma_{11} \; = \; \dfrac{3}{8}+\dfrac{3}{2}Z,
\no \\
\Delta_8 &=& -\Sigma_{12} \; = \; - \dfrac{1}{4}. \label{DELTAS}
\eeqa
We have also checked the values given in (\ref{DELTAS}) by
applying them to the $P_{\ell 3}$ form factors
$f_+^{K^+\pi^0}$, $f_-^{K^+\pi^0}$, $f_-^{K^0 \pi^-}$ and
$f_{\pm}^{\eta \pi}$.

\section{Applications of the Electromagnetic Lagrangian}
\label{sec: Applications}
\renewcommand{\theequation}{\arabic{section}.\arabic{equation}}
\setcounter{equation}{0}

With the methods described in the previous sections, the
electromagnetic contributions of order $e^2 p^2$ to any mesonic
observable can be calculated. So far, only a few results of this
kind have been worked out completely. In Ref. \cite{Urech}, the
diagonal elements of the pseudoscalar mass matrix have been
calculated to $\cO (e^2 p^2)$. The remaining off--diagonal term
related to $\pi^0-\eta$ mixing can be found in Ref.
\cite{NeuRup}. In the same paper, also the $\cO (e^2 p^2)$
contributions to the ratio of $K_{\ell 3}$ form factors
$f_+^{K^+\pi^0}(0) / f_+^{K^0 \pi^-}(0)$ and to the $\eta_{\ell
3}$ form factors $f_{\pm}^{\eta\pi}(t)$ have been given.

For a complete numerical analysis of these results, some information
about the electromagnetic low--energy constants $S_i^r(\mu)$ is
needed. Unfortunately, our present knowledge of these parameters
is restricted to crude order of magnitude estimates. This is in
sharp contrast to the $\cO (p^4)$ coupling constants
$L_i^r(\mu)$ associated with the effective Lagrangian of pure
QCD which have been determined \cite{DAPHNE2} rather accurately
by using experimental input and large $N_c$ arguments \cite{GL2}.
But even with our limited knowledge about the couplings of the $\cO (e^2 p^2)$
Lagrangian, non--trivial results about the possible size of
the electromagnetic contributions can be obtained.

This can be seen, for instance, in the mass spectrum of the pseudoscalars:
The ``magic'' combination \cite{GL2} of kaon and pion masses
\beq
(M^2_{K^0} - M^2_{K^\pm} + M^2_{\pi^\pm} - M^2_{\pi^0})
\cdot \frac{M_\pi^2}{(M_K^2 - M^2_\pi)M_K^2},
\eeq
can be expressed through the masses of the three light quarks and an
electromagnetic term of $\cO (e^2 p^2)$:
\beqa
\frac{m^2_d - m_u^2}{m^2_s - \wh m^2} &=& \left[
(M^2_{K^0} - M^2_{K^\pm} + M^2_{\pi^\pm} - M^2_{\pi^0})_{\rm exp}
- (M^2_{K^0} - M^2_{K^\pm} + M^2_{\pi^\pm} - M^2_{\pi^0})_{\rm
EM}\right] \no \\
&& \cdot \frac{M_\pi^2}{(M_K^2 - M^2_\pi)M_K^2}.  \label{mq-ratio}
\eeqa
The purely electromagnetic quantity \cite{Urech,NeuRup}
\beqa
\label{EM}
(M^2_{K^0} - M^2_{K^\pm} + M^2_{\pi^\pm} - M^2_{\pi^0})_{\rm EM} &=&
 e^2 M_K^2 \left[ \frac{1}{(4\pi)^2}
\left( 3 \ln \frac{M_K^2}{\mu^2} - 4 + 2 Z \ln \frac{M^2_K}{\mu^2}\right)
\right. \no \\
&& \mbox{} +\left. \frac{4}{3} S_2^r(\mu) - 8 S_7^r(\mu) + 16 Z L_5^r(\mu)
\right] + \cO(e^2M_\pi^2) \no \\
\eeqa
gives the deviation from Dashen's limit \cite{Dashen}. The
unknown combination of low--energy constants $S_2^r(\mu) - 6
S_7^r(\mu)$ determines the size of this deviation. Chiral
dimensional analysis \cite{Weinberg,Manohar} suggests the upper bound
\beq
|S_i^r(M_\rho)| \; \lets \; \frac{1}{(4\pi)^2} = 6.3 \cdot 10^{-3}
\label{BOUND}
\eeq
for the coupling constants of the effective Lagrangian.
The resulting bounds
\beq
-\frac{7}{(4\pi)^2} \leq S_2^r(M_\rho) - 6S_7^r(M_\rho) \leq
\frac{7}{(4\pi)^2}. \label{VARIATION}
\eeq
imply the range
\beq
1.5 \cdot 10^{-3} \; \lets \;
1/Q^2 := \frac{m_d^2 - m_u^2}{m_s^2 - \wh m^2}
\; \lets \; 2.4 \cdot 10^{-3}. \label{Qrange}
\eeq
for the combination of quark masses occuring in
(\ref{mq-ratio}). This estimate has to be compared with the value
for $1/Q^2$ in Dashen's limit (corresponding to a vanishing
electromagnetic contribution (\ref{EM})):
\beq
1/Q^2|_{\rm Dashen} = 1.72 \cdot 10^{-3}.
\eeq
Values for $1/Q^2$ rather close to the upper bound of
(\ref{Qrange}) have been obtained by certain model calculations
\cite{DHW,Bijnens} which might also be supported by the present
experimental data on $\eta \ra 3 \pi$ decays.

The size of the parameter $Q$ constitutes an important
ingredient for the determination of
$m_u/m_d$ and $m_s/m_d$ \cite{Leumass}. The potentially
large deviation of $Q$ from its value in the Dashen limit led to
some doubts \cite{Goldman} about the validity of the standard results
\cite{Leumass} for these quark mass ratios. However, taking into
account also the additional constraints from the mass splitting of the
baryons \cite{Gasser,GL6} and from an analysis of $\eta-\eta'$
mixing \cite{GL2}, the possible effects \cite{NeuRup,Brioni} on the
determination of $m_u/m_d$ and $m_s/m_d$ are not too dramatic.

\section{$P_{\ell 2}$ Form Factors}
\label{sec: Pl2}
\renewcommand{\theequation}{\arabic{section}.\arabic{equation}}
\setcounter{equation}{0}

In this section we investigate the contributions of order $e^2
p^2$ to the $P_{\ell 2}$ form factors $F_{\alpha}(X)$. These
quantities are defined by the hadronic matrix elements
\beq
\langle 0 |\bar q(0) \gamma^\mu \gamma_5 X^\dg q(0) |\alpha,p\rangle =
i \sqrt{2}  p^\mu F_{\alpha}(X),
\label{formfactors}
\eeq
where we have used a covariant normalization of one--particle states,
\beq
\langle p'|p\rangle = (2\pi)^3 \; 2p^0 \; \delta^{(3)}
(\vec p{\,}' - \vec p\,).
\eeq
The index $\alpha$ denotes a pseudoscalar mass eigenstate and
the $3 \times 3$ matrix $X$ picks out the desired component
of the axial vector current. For the form factors associated
with the non--vanishing matrix elements we find the following
expressions\footnote{See also Ref. \cite{GL2} for the results in the
limit $e = 0$.}:
\beqa
F_{\pi^\pm} &:=& F_{\pi^+} \left(\frac{\lambda_1 + i \lambda_2}{2}\right)
= F_{\pi^-} \left(\frac{\lambda_1 - i \lambda_2}{2} \right) \no \\
&=& F \left\{ 1 + \frac{4}{F^2} [L_4^r(\mu) (M^2_\pi + 2M_K^2) +
L_5^r (\mu) M^2_\pi] \right. \no \\
&& \mbox{} - \frac{1}{4(4\pi)^2 F^2} \left[ 2M^2_{\pi^\pm} \ln
\frac{M^2_{\pi^\pm}}{\mu^2} + 2 M^2_{\pi^0} \ln \frac{M^2_{\pi^0}}{\mu^2}
+ M^2_{K^\pm} \ln \frac{M^2_{K^\pm}}{\mu^2} + M^2_{K^0} \ln
\frac{M^2_{K^0}}{\mu^2} \right] \no \\
&&\mbox{} + \left.\frac{2e^2}{9} [6S_1^r(\mu) + 5S_2^r(\mu) - 9 S_8^r(\mu)]
+ \frac{e^2}{2(4\pi)^2} \left[ 3 \ln \frac{M^2_\pi}{\mu^2} - 6 -
2 \ln \frac{m^2_\gamma}{\mu^2} \right] \right\}, \label{FORM1}
\eeqa
\beqa
F_{K^\pm} &:=& F_{K^+} \left(\frac{\lambda_4 + i \lambda_5}{2}\right) =
F_{K^-} \left(\frac{\lambda_4 - i \lambda_5}{2} \right)  \no \\
&=& F \left\{ 1 + \frac{4}{F^2} [L_4^r(\mu) (M^2_\pi + 2M_K^2) +
L_5^r (\mu) M^2_K] \right. \no \\
&& \mbox{} - \frac{1}{8(4\pi)^2 F^2} \left[ 2M^2_{\pi^\pm} \ln
\frac{M^2_{\pi^\pm}}{\mu^2} +  M^2_{\pi^0} \ln \frac{M^2_{\pi^0}}{\mu^2}
\right.\no \\
&&\mbox{} + \left. 4M^2_{K^\pm} \ln \frac{M^2_{K^\pm}}{\mu^2} + 2M^2_{K^0} \ln
\frac{M^2_{K^0}}{\mu^2} + 3M^2_\eta \ln \frac{M^2_\eta}{\mu^2} \right] \no \\
&&\mbox{} + \frac{8\sqrt{3}\;\ve}{3F^2} \; L_5^r(\mu) (M^2_\pi - M^2_K)
- \frac{\sqrt{3}\;\ve}{4(4\pi)^2F^2}
\left[ M^2_\pi \ln \frac{M^2_\pi}{\mu^2} - M^2_\eta \ln
\frac{M^2_\eta}{\mu^2} \right] \no \\
&& \mbox{} + \left. \frac{2e^2}{9} [6S_1^r(\mu) + 5S_2^r(\mu) - 9 S_8^r(\mu)]
+ \frac{e^2}{2(4\pi)^2} \left[ 3 \ln \frac{M^2_K}{\mu^2} - 6 -
2 \ln \frac{m^2_\gamma}{\mu^2} \right] \right\}, \label{FORM2}
\eeqa
\beqa
F_{K^0} &:=& F_{K^0} \left(\frac{\lambda_6 + i \lambda_7}{2}\right) =
F_{\bar K^0} \left(\frac{\lambda_6 - i \lambda_7}{2} \right) \no \\
&=& F \left\{ 1 + \frac{4}{F^2} [L_4^r(\mu) (M^2_\pi + 2M_K^2) +
L_5^r (\mu) M^2_K] \right. \no \\
&& \mbox{} - \frac{1}{8(4\pi)^2 F^2} \left[ 2M^2_{\pi^\pm} \ln
\frac{M^2_{\pi^\pm}}{\mu^2} +  M^2_{\pi^0} \ln \frac{M^2_{\pi^0}}{\mu^2}
\right. \no \\
&& \mbox{} + \left. 2M^2_{K^\pm} \ln \frac{M^2_{K^\pm}}{\mu^2} +
4M^2_{K^0} \ln
\frac{M^2_{K^0}}{\mu^2} + 3M^2_\eta \ln \frac{M^2_\eta}{\mu^2} \right] \no \\
&&\mbox{} - \frac{8\sqrt{3}\;\ve}{3F^2} \; L_5^r(\mu) (M^2_\pi - M^2_K)
+ \frac{\sqrt{3}\;\ve}{4(4\pi)^2F^2}
\left[ M^2_\pi \ln \frac{M^2_\pi}{\mu^2} - M^2_\eta \ln
\frac{M^2_\eta}{\mu^2} \right] \no \\
&& \mbox{} + \left.
 \frac{4e^2}{9} [ 3 S_1^r(\mu) + S_2^r(\mu)] \right\}, \label{FORM3}
\eeqa
\beqa
F_{\pi^0}\left( \frac{\lambda_3}{\sqrt{2}}\right)
&=& F \left\{ 1 + \frac{4}{F^2} [L_4^r(\mu) (M^2_\pi + 2M_K^2) +
L_5^r (\mu) M^2_\pi] \right. \no \\
&& \mbox{} - \frac{1}{4(4\pi)^2 F^2} \left[ 4M^2_{\pi^\pm} \ln
\frac{M^2_{\pi^\pm}}{\mu^2} +
 M^2_{K^\pm} \ln \frac{M^2_{K^\pm}}{\mu^2} + M^2_{K^0} \ln
\frac{M^2_{K^0}}{\mu^2} \right] \no \\
&&\mbox{} + \left.\frac{e^2}{9} [12S_1^r(\mu) + 10S_2^r(\mu) + 9 S_3^r(\mu)]
\right\}, \label{FORM4}
\eeqa
\beqa
F_{\pi^0} \left( \frac{\lambda_8}{\sqrt{2}}\right) &=&
F \left\{ \ve - \frac{M^2_{\hat \pi^0 \hat \eta}}{M^2_\eta - M^2_\pi}
+ \frac{4\ve}{F^2} [L_4^r(\mu)(M^2_\pi + 2M^2_K) + L_5^r(\mu)M^2_\pi]\right.
\no \\
&& \mbox{} - \frac{\ve}{2(4\pi)^2F^2} \left[2(M^2_\pi - M^2_K) +
(2M^2_\pi + M^2_K) \ln \frac{M^2_K}{\mu^2} \right] \no \\
&& \mbox{} + \left. \frac{\sqrt{3}\; e^2}{9} [2 S_2^r(\mu) + 3 S_3^r(\mu)]
- \frac{\sqrt{3}\; e^2}{2(4\pi)^2} Z \left[ 1 + \ln \frac{M^2_K}{\mu^2}
\right] \right\}, \label{FORM5}
\eeqa
\beqa
F_\eta \left( \frac{\lambda_3}{\sqrt{2}}\right) &=&
F \left\{- \ve + \frac{M^2_{\hat \pi^0 \hat \eta}}{M^2_\eta - M^2_\pi}
- \frac{4\ve}{F^2} [L_4^r(\mu)(M^2_\pi + 2M^2_K) + L_5^r(\mu)M^2_\eta]\right.
\no \\
&& \mbox{} - \frac{\ve}{2(4\pi)^2F^2} \left[2(M^2_\pi - M^2_K)
- 2M^2_\pi \ln \frac{M^2_\pi}{\mu^2} +
(2M^2_\pi - 3 M^2_K) \ln \frac{M^2_K}{\mu^2} \right] \no \\
&& \mbox{} + \left. \frac{\sqrt{3}\; e^2}{9} [2 S_2^r(\mu) + 3 S_3^r(\mu)]
- \frac{\sqrt{3}\; e^2}{2(4\pi)^2} Z \left[ 1 + \ln \frac{M^2_K}{\mu^2}
\right] \right\}, \label{FORM6}
\eeqa
\beqa
F_\eta \left( \frac{\lambda_8}{\sqrt{2}}\right)
&=& F \left\{ 1 + \frac{4}{F^2} [L_4^r(\mu) (M^2_\pi + 2M_K^2) +
L_5^r (\mu) M^2_\eta] \right. \no \\
&& \mbox{} - \frac{3}{4(4\pi)^2 F^2} \left[
 M^2_{K^\pm} \ln \frac{M^2_{K^\pm}}{\mu^2} + M^2_{K^0} \ln
\frac{M^2_{K^0}}{\mu^2} \right] \no \\
&&\mbox{} + \left.\frac{e^2}{3} [4S_1^r(\mu) + 2S_2^r(\mu) + S_3^r(\mu)]
\right\}. \label{FORM7}
\eeqa
The quantity $M^2_{\hat \pi^0 \hat \eta}$ in (\ref{FORM5}) and (\ref{FORM6})
is the off--diagonal element of the $\pi^0 -
\eta$ mass matrix in the basis of the tree--level mass eigenfields
$\wh \pi^0, \wh \eta$. Its explicit form can be found in Ref. \cite{NeuRup}.
In all our formulas, terms of higher than linear order in the
isospin breaking parameters $\ve, e^2$ have been neglected. The
electromagnetic infrared divergence occuring in
(\ref{FORM1}) and (\ref{FORM2}) has been taken into
account by introducing the small photon mass $m_{\gamma}$. Taken
for themselves, the expressions given above are not observable
quantities but only (major) parts in a full analysis of $P_{\ell 2}$
decays. The infrared divergences are absorbed by adding the
corresponding $P_{\ell 2 \gamma}$ contributions \cite{BEG}. Furthermore,
the leptonic part together with the associated electromagnetic
corrections has to be included \cite{XXX}. However, for our
present purposes, the information contained in (\ref{FORM1}--\ref{FORM7})
will be sufficient.

To get a feeling for the possible size of the electromagnetic
contributions to isospin violating quantities we build the ratio
\beqa
R &:=& \frac{F_{K^0} F_{\pi^\pm} }{F_{K^\pm} F_{\pi^0}
(\lambda_3/ \sqrt{2})}
= 1 + \frac{4\ve}{\sqrt{3}} \left\{
\frac{F_K}{F_\pi} - 1 + \frac{1}{4(4\pi)^2F^2} \left[ M^2_\pi - M^2_K +
M^2_\pi \ln \frac{M^2_K}{M^2_\pi} \right] \right\} \no \\
&& \mbox{} - \frac{e^2}{3} [2S_2^r(\mu) + 3 S_3^r(\mu)]
+ \frac{3e^2}{2(4\pi)^2} \left[ Z \left( 1 + \ln \frac{M^2_K}{\mu^2}
\right) - \ln \frac{M^2_K}{M^2_\pi} \right].
\label{R}
\eeqa
In this specific combination of the form factors (\ref{FORM1}--\ref{FORM4}),
the infrared divergent terms cancel. The only remaining
uncertainty in (\ref{R}) is the electromagnetic low--energy constant
$2 S_2^r(\mu) + 3 S_3^r(\mu)$. At this point, we completely
disregard the question if it will ever be possible to determine
the quantity R with a sufficient experimental accuracy.
We just want to compare the size of the electromagnetic and the QCD
part contained in (\ref{R}).
With
\beq
\ve = (1.00 \pm 0.07) \cdot 10^{-2}, \label{numeps}
\eeq
extracted from the mass splitting in the baryon octet
\cite{Leumass,Gasser,GL6} and $F_K/F_{\pi} = 1.22$ \cite{PDG},
we find
\beq
(R - 1)_{\rm QCD} = 4.4 \cdot 10^{-3}.
\label{RQCD}
\eeq
Assuming the validity of (\ref{BOUND}), we expect an electromagnetic
contribution within the range
\beq
-3.2 \cdot 10^{-3}\; \lets \; (R - 1)_{\rm EM} \; \lets \; -1.2 \cdot 10^{-3},
\label{REM}
\eeq
where the lower (upper) bound corresponds to $2 S_2^r(M_{\rho}) +
3 S_3^r(M_{\rho}) = {}^{\;\,+}_{(-)} 5 / (4 \pi)^2$.
We conclude from
(\ref{RQCD}) and (\ref{REM}) that, in general, isospin violating
terms of electromagnetic origin can be of equal importance as
the corresponding QCD pieces proportional to the quark mass
difference $m_d - m_u$.

\section{$K_{\ell 3}$ Form Factors}
\label{sec: Kl3}
\renewcommand{\theequation}{\arabic{section}.\arabic{equation}}
\setcounter{equation}{0}

Finally, we discuss the $K_{\ell 3}$ form factors
$f_+^{K^+ \pi^0}(0)$ and $f_+^{K^0 \pi^-}(0)$ including the
electromagnetic contributions of $\cO(e^2 p^2)$. Our results are
given by
\beqa
f_+^{K^+\pi^0}(0) &=& 1 + \frac{1}{2} H_{K^\pm \pi^0}(0) + \frac{3}{2}
H_{K^\pm \eta}(0) + H_{K^0 \pi^\pm}(0) \no \\
&& \mbox{} + \sqrt{3} \left(\ve - \frac{M^2_{\hat \pi^0 \hat \eta}}
{M^2_\eta -
M^2_\pi} \right) + \sqrt{3} \; \ve \left[ \frac{5}{2} H_{K\pi}(0) +
\frac{1}{2} H_{K\eta}(0)\right] \no \\
&& \mbox{} - \frac{e^2}{(4\pi)^2} \left[ 2 + \ln \frac{m^2_\gamma}{M^2_K}
+ \frac{1}{4} \ln \frac{M^2_K}{\mu^2} + 2(4\pi)^2 S_8^r(\mu)\right],
\label{Kplus/pi0}
\eeqa
and
\beqa
f_+^{K^0\pi^-}(0) &=& 1 +  H_{K^0 \pi^\pm}(0) + \frac{1}{2}
H_{K^\pm \pi^0}(0) + \frac{3}{2} H_{K^\pm \eta}(0) \no \\
&& \mbox{} +
 \sqrt{3} \; \ve \left[  H_{K\pi}(0) -
 H_{K\eta}(0)\right] \no \\
&& \mbox{} - \frac{e^2}{(4\pi)^2} \left[ 2 + \ln \frac{m^2_\gamma}{M^2_{\pi}}
+ \frac{1}{4} \ln \frac{M^2_{\pi}}{\mu^2} + 2(4\pi)^2 S_8^r(\mu)\right].
\label{K0/piminus}
\eeqa
The function $H_{PQ}(t)$ was defined in \cite{GL5}, where also
$f_+^{K^+ \pi^0}(t)$,  $f_+^{K^0 \pi^-}(t)$ in the limit $e = 0$ were
presented.

In the ratio
\beq
r_{K\pi} = \frac{f_+^{K^+\pi^0}(0)}{f_+^{K^0 \pi^-}(0)} =
1 + \sqrt{3} \left( \ve - \frac{M^2_{\hat \pi^0 \hat \eta}}
{M_\eta^2 - M_\pi^2} \right) + \frac{3e^2}{4(4\pi)^2} \ln
\frac{M_K^2}{M_\pi^2},
\label{RKPI}
\eeq
the infrared divergent terms cancel. As in the previous example (\ref{R}),
only $2 S_2^r(\mu) + 3 S_3^r(\mu)$ (contained in $M^2_{\hat \pi^0 \hat \eta}$)
remains as an unknown parameter. Disentangling the QCD and the electromagnetic
contribution to (\ref{RKPI}) one finds \cite{NeuRup}
\beq
(r_{K \pi} - 1)_{\rm QCD} = 2.1 \cdot 10^{-2},
\label{RKPIQCD}
\eeq
and
\beq
0 \; \lets \; (r_{K \pi} - 1)_{\rm EM} \; \lets \; 0.2 \cdot 10^{-2},
\label{RKPIEM}
\eeq
respectively, where (\ref{RKPIEM}) is again based on
(\ref{BOUND}). In spite of our ignorance of the exact values of
the electromagnetic coupling constants, we have obtained a
rather precise result: The electromagnetic contribution to
$r_{K \pi} - 1$ can increase the pure QCD value by at most 10 \%.

Let us also compare the theoretical results (\ref{RKPIQCD})
and (\ref{RKPIEM}) with the present experimental data.
Dividing the rates of $K^+ \ra \pi^0 e^+ \nu_e$ and
$K^0 \ra \pi^- e^+ \nu_e$ by the relevant phase space integrals
(including those
electromagnetic corrections which are sensitive to the lepton
kinematics \cite{LeuRoos}) one finds \cite{GL5}
\beq
|\frac{f_+^{K^+\pi^0}(0)}{f_+^{K^0 \pi^-}(0)}|^2 = 1.057 \pm 0.019,
\label{EXPRAT}
\eeq
which implies
\beq
(r_{K \pi} - 1)_{\rm exp} = (2.8 \pm 0.9) \cdot 10^{-2}.
\label{RKPIEXP}
\eeq
This means that the error in the present data is still much larger than the
theoretical uncertainty due to electromagnetism.

\section{Conclusions}
\label{sec: Conclusions}
\renewcommand{\theequation}{\arabic{section}.\arabic{equation}}
\setcounter{equation}{0}

We have used the machinery of chiral perturbation theory
including a systematic treatment of the electromagnetic
interaction \cite{Urech}. Within this theoretical framework, a
one--loop--analysis allows the computation of the pure QCD
contributions to $\cO(p^4)$ and of the electromagnetic part to
$\cO(e^2 p^2)$ for any observable in the sector of pseudoscalar
mesons. The low--energy constants associated with the $\cO(p^4)$
effective Lagrangian of strong interactions are well known
parameters. For the coupling constants of the $\cO(e^2 p^2)$
electromagnetic Lagrangian, only order of magnitude estimates
based on chiral dimensional analysis are presently available.

This situation might change by future precision measurements of
isospin breaking observables or, on the theoretical side, by a
determination of the relevant low--energy constants using chiral
models or even lattice calculations (for examples in the strong
sector see Ref. \cite{Ecker} and the citations therein). Such an
improvement of our knowledge about the $\cO(e^2 p^2)$ coupling
constants would also drastically increase the value of
our formal expressions for the electromagnetic
contributions to several isospin breaking quantities.

We have performed a one--loop analysis of all
$P_{\ell 2}$ and the $K_{\ell 3}$ form factors $f_+^{K^+
\pi^0}(0)$ and $f_+^{K^0 \pi^-}(0)$. Our results allow a
discussion of the magnitude of isospin violating effects due to
pure QCD, that is the difference of the up and down quark
masses, and those originating from QED isospin violation. There
is no general feature, the size of the respective contributions
depends strongly on the observed quantity.

For example, in the specific combination of $P_{\ell 2}$ form
factors $R - 1$ (defined in (\ref{R})), the isospin violating
effects of electromagnetic origin can be of equal size as the
QCD ones. Similarly, it has been found \cite{Urech,NeuRup} that
sizable deviations from Dashen's limit \cite{Dashen} for the
pseudoscalar meson masses cannot be excluded. On the other hand,
for the ratio $r_{K\pi} = f_+^{K^+\pi^0}(0) / f_+^{K^0
\pi^-}(0)$ of $K_{\ell 3}$ form factors we have obtained the
rather precise result that the electromagnetic contribution to
$r_{K\pi} - 1$ can at most be 10 \% of the corresponding QCD
value, which is quite similar in the case of the $\eta_{\ell 3}$
form factors $f_{\pm}^{\eta\pi}(t)$ \cite{NeuRup}.

At present, the experimental errors are still much larger
than the uncertainties induced by electromagnetic isospin
violating contributions. But our analysis shows quite clearly
that if isospin violating effects due to $m_u \not= m_d$ are
considered with one--loop accuracy, one also has to take into account
electromagnetic effects up to $\cO(e^2 p^2)$.

\subsection*{Acknowledgements}

We thank G. Ecker for reading the manuscript.

\end{document}